\pdfoutput=1
\PassOptionsToPackage{pdfpagelabels=false}{hyperref}
\documentclass[fleqn,usenatbib,useAMS]{mnras}
\usepackage[british]{babel}             

\usepackage{newtxtext,newtxmath}

\usepackage[T1]{fontenc}

\usepackage{graphicx}    
\usepackage{amsmath}    
\usepackage{amssymb}    
\usepackage{CJK}
\usepackage{etoolbox}
\makeatletter
\patchcmd\@combinedblfloats{\box\@outputbox}{\unvbox\@outputbox}{}{%
   \errmessage{\noexpand\@combinedblfloats could not be patched}%
}%
 \makeatother
\newcommand{\feh}{\ensuremath{[\textrm{Fe}/\textrm{H}]}}

\newcommand{\teff}{\ensuremath{T_\textrm{eff}}}
\newcommand{\logg}{\ensuremath{\log \textrm{g}}}
\defcitealias{Oh2016}{O17}	

\begin{document}
\begin{CJK*}{UTF8}{gbsn}
\label{firstpage}
\pagerange{\pageref{firstpage}--\pageref{lastpage}}


\title[GALAH Survey: Co-orbiting stars]{The GALAH survey: Co-orbiting stars and chemical tagging}

\author[J. D. Simpson et al.]{Jeffrey D. Simpson$^{1}$\thanks{Email: \texttt{jeffrey.simpson@aao.gov.au}},
Sarah~L.~Martell$^{2,18}$\thanks{Email: \texttt{s.martell@unsw.edu.au}},
Gary~Da~Costa$^{3}$,
Andrew~R.~Casey$^{4,5}$\newauthor
Ken~C.~Freeman$^{3}$,
Jonathan~Horner$^{6}$,
Yuan-Sen~Ting$^{7,8,9}$,
David~M.~Nataf$^{10}$,\newauthor
Geraint~F.~Lewis$^{11}$,
Melissa~K.~Ness$^{12,13}$,
Daniel~B.~Zucker$^{1,14}$,
Peter~L.~Cottrell$^{15,16}$,\newauthor
Klemen~\v{C}otar$^{17}$,
Martin~Asplund$^{3,18}$,
Joss~Bland-Hawthorn$^{4,11}$,
Sven Buder$^{19}$,\newauthor
Valentina~{D'Orazi}$^{20}$,
Gayandhi~M.~De~Silva$^{1,11}$,
Ly~Duong$^{3,18}$,
Janez~Kos$^{11}$,\newauthor
Jane~Lin$^{3,18}$,
Karin~Lind$^{19,21}$,
Katharine~J.~Schlesinger$^{3}$,
Sanjib~Sharma$^{11}$,\newauthor
Toma\v{z}~Zwitter$^{17}$,
Prajwal~R.~Kafle$^{22}$,
Thomas~Nordlander$^{3,18}$
\\
$^{1}$Australian Astronomical Observatory, 105 Delhi Rd, North Ryde, NSW 2113, Australia\\
$^{2}$School of Physics, UNSW, Sydney, NSW 2052, Australia\\
$^{3}$Research School of Astronomy \& Astrophysics, Australian National University, ACT 2611, Australia\\
$^{4}$School of Physics and Astronomy, Monash University, Clayton 3800, Victoria, Australia\\
$^{5}$Faculty of Information Technology, Monash University, Clayton 3800, Victoria, Australia\\
$^{6}$University of Southern Queensland, Toowoomba, Queensland 4350, Australia\\
$^{7}$Institute for Advanced Study, Princeton, NJ 08540, USA \\
$^{8}$Department of Astrophysical Sciences, Princeton University, Princeton, NJ 08544, USA \\
$^{9}$Observatories of the Carnegie Institution of Washington, 813 Santa Barbara Street, Pasadena, CA 91101, USA \\
$^{10}$Center for Astrophysical Sciences and Department of Physics and Astronomy, The Johns Hopkins University, Baltimore, MD 21218, USA\\
$^{11}$Sydney Institute for Astronomy, School of Physics, A28, The University of Sydney, NSW, 2006, Australia\\
$^{12}$Department of Astronomy, Columbia University, Pupin Physics Laboratories, New York, NY 10027, USA\\
$^{13}$Center for Computational Astrophysics, Flatiron Institute, 162 Fifth Avenue, New York, NY 10010, USA \\
$^{14}$Department of Physics and Astronomy, Macquarie University, Sydney, NSW 2109, Australia \\
$^{15}$School of Physical and Chemical Sciences, University of Canterbury, New Zealand\\
$^{16}$Monash Centre for Astrophysics, School of Physics and Astronomy, Monash University, Australia\\
$^{17}$Faculty of Mathematics and Physics, University of Ljubljana, Jadranska 19, 1000 Ljubljana, Slovenia\\
$^{18}$Centre of Excellence for Astrophysics in Three Dimensions (ASTRO-3D), Australia\\
$^{19}$Max Planck Institute for Astronomy (MPIA), Koenigstuhl 17, 69117 Heidelberg, Germany\\
$^{20}$Istituto Nazionale di Astrofisica, Osservatorio Astronomico di Padova, vicolo dell'Osservatorio 5, 35122, Padova, Italy \\
$^{21}$Department of Physics and Astronomy, Uppsala University, Box 517, SE-751 20 Uppsala, Sweden\\
$^{22}$ICRAR, The University of Western Australia, 35 Stirling Highway, Crawley, WA 6009, Australia \\
}

\date{Accepted XXX. Received YYY; in original form ZZZ}

\pubyear{2018}


\maketitle
\end{CJK*}

\begin{abstract}
We present a study using the second data release of the GALAH survey of stellar parameters and elemental abundances of 15 pairs of stars identified by \citet{Oh2016}. They identified these pairs as potentially co-moving pairs using proper motions and parallaxes from \textit{Gaia} DR1. We find that 11 very wide ($>1.7$~pc) pairs of stars do in fact have similar Galactic orbits, while a further four claimed co-moving pairs are not truly co-orbiting. Eight of the 11 co-orbiting pairs have reliable stellar parameters and abundances, and we find that three of those are quite similar in their abundance patterns, while five have significant \feh\ differences. For the latter, this indicates that they could be co-orbiting because of the general dynamical coldness of the thin disc, or perhaps resonances induced by the Galaxy, rather than a shared formation site. Stars such as these, wide binaries, debris of past star formation episodes, and coincidental co-orbiters, are crucial for exploring the limits of chemical tagging in the Milky Way.
\end{abstract}

\begin{keywords}
stars: abundances, Galaxy: disc, stars: formation
\end{keywords}



\section{Introduction}\label{sec:introduction}
The GALactic Archaeology with HERMES (GALAH) survey is a large and ambitious spectroscopic investigation of the local stellar environment \citep{DeSilva2015a}. One of its principal aims is to determine precise abundances of nearly 30 elements\footnote{While measurements of nearly 30 elements are possible from spectra obtained with HERMES, in GALAH DR2 we report abundances for 23 elements, and in this work consider only the abundances of 19 elements that were present in our stars of interest.} from one million stars and to use chemical tagging to identify dispersed stellar clusters in the field of the disc and halo \citep[for the initial motivating papers, see][]{Freeman2002, Bland-Hawthorn2010}. It relies on the assumption that although they may disperse into different regions of kinematic phase space, the stars that form within a single cluster will continue to possess a common and unique pattern of chemical abundances. Chemically tagging the stars from many formation sites would enable us to unravel the formation and evolutionary history of the Galaxy in a way that it is not possible from their spatial, photometric, or kinematic properties.

Chemical tagging solely in abundance space is a challenging task, and there is much discussion in the literature about the prospects of the technique being successful \citep [e.g.,][]{Ting2015, Bovy2016, Hogg2016}. The ultimate goal of GALAH is to perform this chemical tagging without the recourse to other information --- i.e., kinematics --- and so cases in which we can test for coherence in both kinematics and chemical composition are an important step toward that goal. Stars in streams and moving groups fall between the extremes of stars still in their formation clusters and the majority of disc stars that have entirely lost that original spatial and kinematic coherence. They are a critical test set for chemical tagging since their orbital similarities can provide a confirmation of the shared formation history which we would infer from their compositions.

Spectra obtained in the GALAH survey provide the radial velocities of the stars, but we require full 6D (position, velocity) phase space information about the stars to place these stars in streams and moving groups. The ESA \textit{Gaia} mission \citep{Prusti2016} provides us with this. The first data release \citep{Brown2016} was utilized by a number of authors to identify potentially co-moving pairs and groups of stars \citep{Oh2016, Andrews2017, Oelkers2017}. These studies have each adopted different methods and goals for their searches. The work for this paper was primarily performed prior to the release of \textit{Gaia} DR2 \citep{Brown2018}, but we use its parallaxes and proper motions. 

We will focus on the pairs of stars identified by \citet[][hereafter \citetalias{Oh2016}]{Oh2016}. These stars are all found within $\sim600$~pc of the Sun, which is where the errors in parallax found by the Tycho-\textit{Gaia} astrometric solution \citep[TGAS;][]{Michalik2014, Lindegren2016} are small enough\footnote{Using the \citetalias{Oh2016} definition of requiring the parallax signal-to-noise ratio $\varpi/\sigma_\varpi>8$.} to permit a reliable determination of distances and orbits. \citetalias{Oh2016} used the TGAS data to identify over 13000 pairs of co-moving stars with separations less than 10~pc. Because \textit{Gaia} DR1 did not contain radial velocity information for the stars, they had to marginalize over the unknown 3D velocities of the stars. In their method, each star can be paired with multiple other stars, and many of the pairs they identified were parts of larger networks. Their analysis recovered several known clusters, including the Pleiades, the Sco-Cen young stellar association, the Hyades, and NGC~2632. However, most of their groups do not have a known counterpart in the literature, and many were isolated pairs of stars.

Interestingly, by requiring that the proper motions of the stars be highly similar, \citetalias{Oh2016} might reject close binary star systems as potential co-moving pairs. From the calculations of \citet{Andrews2017}, the semi-amplitude of the orbital velocity in the systems would be $>5$~km\,s$^{-1}$ for systems with separations $<15$~AU. If a significant component of this motion were oriented in the plane of the sky, it is easy to imagine that the two stars would appear to have a relative proper motion too large to allow them to orbit together, even though in truth they follow their common barycentre around its orbit.

Further investigations of some of the pairs identified by \citetalias{Oh2016} has been done with low-resolution spectroscopy  \citep{Price-Whelan2017}, infrared photometry \citep{Bochanski2018}, and for one pair, high-resolution spectra \citep[][using results from \citealt{Brewer2016}]{Oh2018}. \textcolor{red}{\citet{Andrews2018} found that their candidate wide binary stars typically had very similar metallicity, using the public catalogues from from the RAVE and LAMOST spectroscopic surveys. However, LAMOST does not publish detailed abundances based on their low-resolution spectra, and the abundance precision of the RAVE catalogue was not high enough to pursue further chemical tagging.} 

\textcolor{red}{Using data from the GALAH survey, we can expand on these studies by adding critical information}. Not only does high-resolution spectroscopy provide radial velocities that allow complete orbital calculations, GALAH also derives stellar parameters and elemental abundances: \teff, \logg, \feh, and abundances for up to 23 elements. For these candidate co-moving pairs, which may be from dissolving and disrupting clusters, we can evaluate whether they have common origins with the aid of kinematic and chemical information.

Spectroscopic stellar parameters and abundances from the GALAH survey make it possible for us to distinguish between different types of moving groups. There are known groups that are `true' moving groups of stars, consisting of the disrupted remnants of old clusters: e.g., HR 1614, Wolf 630 and the Argus moving groups \citep{DeSilva2007, Bubar2010, DeSilva2013}. However, there are other groups \citep[e.g., the Hercules group,][]{Bensby2007, Quillen2018} which have distinctly different chemical abundances and are on similar orbits as a result of dynamical resonances within the Galaxy. Simply relying on kinematics would identify that these are true groups of co-moving stars, but would not provide a full picture of the chemodynamical history of the Galaxy.

This work is structured as follows: data reduction and abundance analysis (Section \ref{sec:data_reduction}); kinematic evaluation of the groups observed by GALAH (Section \ref{sec:groups}); investigation of the abundance patterns of the co-orbiting pairs (Section \ref{sec:abundances}); and a discussion of the intrinsic limits of and future prospects for chemical tagging (Section \ref{sec:discussion}).

\section{Observations and spectrum analysis}\label{sec:data_reduction}

We make use of an internally released catalogue of a similar size and composition to GALAH survey's second data release \citep[GALAH DR2; see the release paper:][]{Buder2018} which maximized the overlap between GALAH and \citetalias{Oh2016}. It is based upon spectra obtained between 2014 January and 2018 January using the 3.9-metre Anglo-Australian Telescope with the HERMES spectrograph \citep{Sheinis2015} and the Two-Degree Field (2dF) top-end \citep{Lewis2002}. 2dF allows for the concurrent acquisition of up to $\sim360$ science targets per exposure. HERMES simultaneously acquires spectra using four independent cameras with non-contiguous wavelength coverage totalling $\sim1000$~\AA\ at a spectral resolving power of $R\approx28,000$. Its fixed wavelength bands are 4715--4900~\AA, 5649--5873~\AA, 6478--6737~\AA, and 7585--7887~\AA. For details on the observing procedures see \citet{Martell2017a} and \citet{Buder2018}. The spectra were reduced using an \textsc{iraf}-based pipeline that was developed specifically for GALAH and optimized for speed, accuracy, and consistency. We direct the reader to \citet{Kos2017} for a detailed description of the reduction procedure.

The GALAH stellar parameter and abundance pipeline description can be found in \citet{Buder2018}. Briefly, the pipeline uses a two-step process. In the first step, spectra with high signal-to-noise are identified and analyzed with the spectrum synthesis code Spectroscopy Made Easy \citep[SME;][]{Valenti1996,Piskunov2017} to determine the stellar labels (\teff, \logg, \feh, $v_\textrm{mic}$, $v\sin i$, $v_\textrm{rad}$, and [X/Fe]). This training set includes the \textit{Gaia} benchmark stars, globular and open cluster stars, and stars with accurate asteroseismic surface gravity from K2 Campaign 1 \citep{Stello2017}. In the second step, \textit{The Cannon} \citep{Ness2015} learns the training set labels from SME and builds a quadratic model at each pixel of the normalized spectrum as a function of the labels. Abundance estimates are then generated from \textit{The Cannon} model.

Overall, the GALAH release used in this work contains a total of 365,516 stars with up to 23 elemental abundances per star. For a minority of stars, the label results from \textit{The Cannon} are not reliable: the label result could be too far from the training set, the $\chi^2$ between the observed spectrum and the spectrum calculated by \textit{The Cannon} could be too large, or the spectra could have been classified by t-SNE \citep[for details on the application of t-SNE to GALAH spectra see][]{Traven2017} as having problems. In addition, the individual elemental abundance can be flagged for similar reasons via the \texttt{flag\_x\_fe}. In this work, we only use abundance values for which \texttt{flag\_x\_fe} is zero, which means that the particular abundance is likely to be reliable.



\section{Co-Moving Groups in GALAH}\label{sec:groups}

\begin{table*}
\centering
\caption{The stellar parameters and orbital characteristics of the 15 pairs. The stellar parameters are given for stars with $\mathrm{\texttt{flag\_cannon}}==0$ (i.e., they are not believed to be unreliable). For each star, 1000 random samples of the 6D information of the stars taking account their uncertainties and covariances were created and then \textsc{galpy} used to integrate the orbit to find the median eccentricity, $z_\mathrm{max}$, perigalacticon, and apogalacticon of the orbit. The uncertainties are the 5th and 95 percentiles of the distributions. The ordering is the same as in Table \ref{table:basic_data}, namely increasing $\Delta(\mathrm{U,V,W})$.}
\label{table:pair_stellar_params}
\begin{tabular}{rrrrrrrrrrrrrrr}
\hline
Group & \texttt{sobject\_id} & \teff\ & \logg & \feh & $G$ & $G_\mathrm{BP}-G_\mathrm{RP}$ & e & $z_\mathrm{max}$& peri& apo\\
 &  & (K) &  &  &  &  & & (pc) &  (kpc) &  (kpc) \\
\hline
3410 & 160813001601030 & $6106\pm55$ & $4.28\pm0.15$ & $-0.47\pm0.07$ & 10.98 & 0.73  & $0.07\pm0.00$ & $29\pm1$ & $7.15\pm0.01$ & $8.31\pm0.01$ \\
3410 & 160813001601029 & $5664\pm73$ & $4.49\pm0.18$ & $-0.36\pm0.08$ & 11.86 & 0.86  & $0.07\pm0.00$ & $28\pm1$ & $7.17\pm0.02$ & $8.32\pm0.01$ \\
\noalign{\vskip 1mm}
3612 & 150211003701379 & $5533\pm67$ & $4.55\pm0.17$ & $+0.23\pm0.08$ & 12.02 & 0.95  & $0.12\pm0.00$ & $87\pm1$ & $6.81\pm0.01$ & $8.70\pm0.02$ \\
3612 & 150211003701380 & $5371\pm61$ & $4.51\pm0.16$ & $+0.23\pm0.07$ & 12.36 & 1.01  & $0.12\pm0.00$ & $89\pm1$ & $6.79\pm0.02$ & $8.66\pm0.03$ \\
\noalign{\vskip 1mm}
3 & 160125004501147 &  &  &  & 10.26 & 1.13  & $0.05\pm0.00$ & $48\pm0$ & $7.12\pm0.02$ & $7.95\pm0.00$ \\
3 & 160130006301220 &  &  &  & 9.79 & 1.01  & $0.06\pm0.00$ & $52\pm1$ & $7.10\pm0.03$ & $7.95\pm0.00$ \\
\noalign{\vskip 1mm}
237 & 170615003401085 &  &  &  & 9.78 & 0.44  & $0.05\pm0.00$ & $123\pm2$ & $7.16\pm0.02$ & $7.89\pm0.01$ \\
237 & 160817001601245 &  &  &  & 9.22 & 0.27  & $0.05\pm0.02$ & $136\pm17$ & $7.21\pm0.25$ & $7.90\pm0.05$ \\
\noalign{\vskip 1mm}
987 & 170516000601281 & $6056\pm51$ & $4.16\pm0.14$ & $-0.31\pm0.06$ & 10.50 & 0.69  & $0.13\pm0.00$ & $284\pm2$ & $6.78\pm0.03$ & $8.89\pm0.03$ \\
987 & 170516000601016 & $6130\pm52$ & $4.18\pm0.14$ & $-0.22\pm0.06$ & 10.49 & 0.68  & $0.12\pm0.00$ & $287\pm10$ & $7.14\pm0.06$ & $9.06\pm0.04$ \\
\noalign{\vskip 1mm}
 
40 & 170711001501145 & $6387\pm18$ & $4.13\pm0.06$ & $-0.19\pm0.02$ & 9.09 & 0.60  & $0.06\pm0.00$ & $57\pm1$ & $7.11\pm0.02$ & $7.95\pm0.01$ \\
40 & 150706001601135 &  &  &  & 9.66 & 1.01  & $0.04\pm0.01$ & $59\pm4$ & $7.41\pm0.16$ & $8.01\pm0.02$ \\
\noalign{\vskip 1mm}
1313 & 170615003401071 & $5568\pm63$ & $4.07\pm0.17$ & $+0.30\pm0.08$ & 11.62 & 0.96  & $0.11\pm0.00$ & $206\pm4$ & $6.42\pm0.03$ & $8.00\pm0.00$ \\
1313 & 160815002101306 & $6219\pm61$ & $4.09\pm0.16$ & $-0.26\pm0.07$ & 10.94 & 0.71  & $0.10\pm0.00$ & $168\pm6$ & $6.44\pm0.03$ & $7.89\pm0.01$ \\
\noalign{\vskip 1mm}
3496 & 160611003101049 & $6137\pm54$ & $4.29\pm0.15$ & $-0.02\pm0.07$ & 9.74 & 0.68  & $0.04\pm0.00$ & $128\pm1$ & $7.36\pm0.02$ & $8.05\pm0.00$ \\
3496 & 160611003101279 & $5190\pm90$ & $4.54\pm0.20$ & $+0.29\pm0.09$ & 11.60 & 1.08  & $0.05\pm0.00$ & $124\pm1$ & $7.15\pm0.02$ & $7.91\pm0.00$ \\
\noalign{\vskip 1mm}
1220 & 150703001601389 & $5733\pm60$ & $4.49\pm0.16$ & $+0.11\pm0.07$ & 10.95 & 0.93  & $0.12\pm0.00$ & $42\pm2$ & $6.79\pm0.02$ & $8.72\pm0.01$ \\
1220 & 170712001601389 & $5988\pm57$ & $4.42\pm0.15$ & $-0.17\pm0.07$ & 10.73 & 0.82  & $0.15\pm0.00$ & $34\pm2$ & $6.65\pm0.04$ & $9.06\pm0.02$ \\
\noalign{\vskip 1mm}
1223 & 170712001601319 & $5848\pm55$ & $4.21\pm0.15$ & $-0.19\pm0.07$ & 10.28 & 0.80  & $0.18\pm0.00$ & $56\pm2$ & $6.41\pm0.03$ & $9.18\pm0.02$ \\
1223 & 150703001601348 & $5791\pm65$ & $4.34\pm0.17$ & $+0.02\pm0.08$ & 10.64 & 0.96  & $0.16\pm0.00$ & $79\pm3$ & $7.16\pm0.05$ & $9.83\pm0.02$ \\
\noalign{\vskip 1mm}
4512 & 161009002601246 & $5770\pm44$ & $4.30\pm0.12$ & $+0.09\pm0.05$ & 12.18 & 0.82  & $0.20\pm0.01$ & $283\pm56$ & $5.64\pm0.06$ & $8.44\pm0.01$ \\
4512 & 161009002601314 & $5599\pm49$ & $4.09\pm0.13$ & $+0.29\pm0.06$ & 11.92 & 0.89  & $0.18\pm0.00$ & $254\pm20$ & $5.99\pm0.04$ & $8.55\pm0.01$ \\
\noalign{\vskip 1mm}
3959 & 170531001901267 & $5899\pm58$ & $4.22\pm0.16$ & $+0.16\pm0.07$ & 11.02 & 0.85  & $0.19\pm0.00$ & $137\pm1$ & $6.23\pm0.03$ & $9.16\pm0.02$ \\
3959 & 160524002101209 & $5814\pm59$ & $4.47\pm0.16$ & $-0.11\pm0.07$ & 11.45 & 0.89  & $0.21\pm0.00$ & $208\pm2$ & $5.60\pm0.03$ & $8.52\pm0.02$ \\
\noalign{\vskip 1mm}
3560 & 170513004901374 & $5293\pm61$ & $4.37\pm0.16$ & $-0.09\pm0.07$ & 11.12 & 1.00  & $0.21\pm0.01$ & $18\pm4$ & $5.60\pm0.10$ & $8.53\pm0.01$ \\
3560 & 170615003901348 & $5138\pm67$ & $4.48\pm0.17$ & $+0.35\pm0.08$ & 11.31 & 1.09  & $0.24\pm0.00$ & $75\pm2$ & $5.69\pm0.02$ & $9.33\pm0.02$ \\
\noalign{\vskip 1mm}
271 & 160423002201186 & $6053\pm43$ & $4.23\pm0.12$ & $-0.04\pm0.05$ & 10.77 & 0.72  & $0.19\pm0.00$ & $73\pm1$ & $5.73\pm0.02$ & $8.41\pm0.01$ \\
271 & 160522002101256 & $5581\pm61$ & $4.36\pm0.16$ & $+0.14\pm0.07$ & 11.64 & 0.93  & $0.13\pm0.00$ & $116\pm3$ & $7.39\pm0.04$ & $9.50\pm0.03$ \\
\noalign{\vskip 1mm}
3027 & 160513001101131 & $5798\pm52$ & $4.50\pm0.14$ & $+0.08\pm0.06$ & 11.25 & 0.85  & $0.09\pm0.00$ & $129\pm1$ & $7.23\pm0.04$ & $8.68\pm0.01$ \\
3027 & 160513001101351 & $5358\pm63$ & $4.52\pm0.17$ & $+0.02\pm0.08$ & 11.84 & 1.00  & $0.14\pm0.00$ & $252\pm1$ & $5.96\pm0.01$ & $7.93\pm0.00$ \\
\noalign{\vskip 1mm}
 
\hline
\end{tabular}
\end{table*}

\citetalias{Oh2016} identified 10,606 stars to be in non-exclusive co-moving pairs or groups\footnote{Group numbers referred to in this work are the \texttt{Group} column of \citetalias{Oh2016}.}. Unfortunately, only 117 of these stars are found in the GALAH catalogue, and only 15 pairs had both stars observed (i.e., 30 stars). Table \ref{table:pair_stellar_params} lists the stellar parameters, photometry and orbital parameters for the 15 pairs. This very small overlap is the result of two selection effects within GALAH. First, the majority of stars in GALAH are found in the magnitude range $12<V<14$, with a smaller number of stars up to $V=9$, while TGAS (used by \citetalias{Oh2016}) is predominantly $G<11$. This means that most of the \citetalias{Oh2016} stars are brighter than GALAH's magnitude range. Second, GALAH only observes stars with $-80^\circ<\delta<+10^\circ$ and $10^\circ<|b|\lesssim50^\circ$. 

\begin{figure*}
    \includegraphics[width=\textwidth]{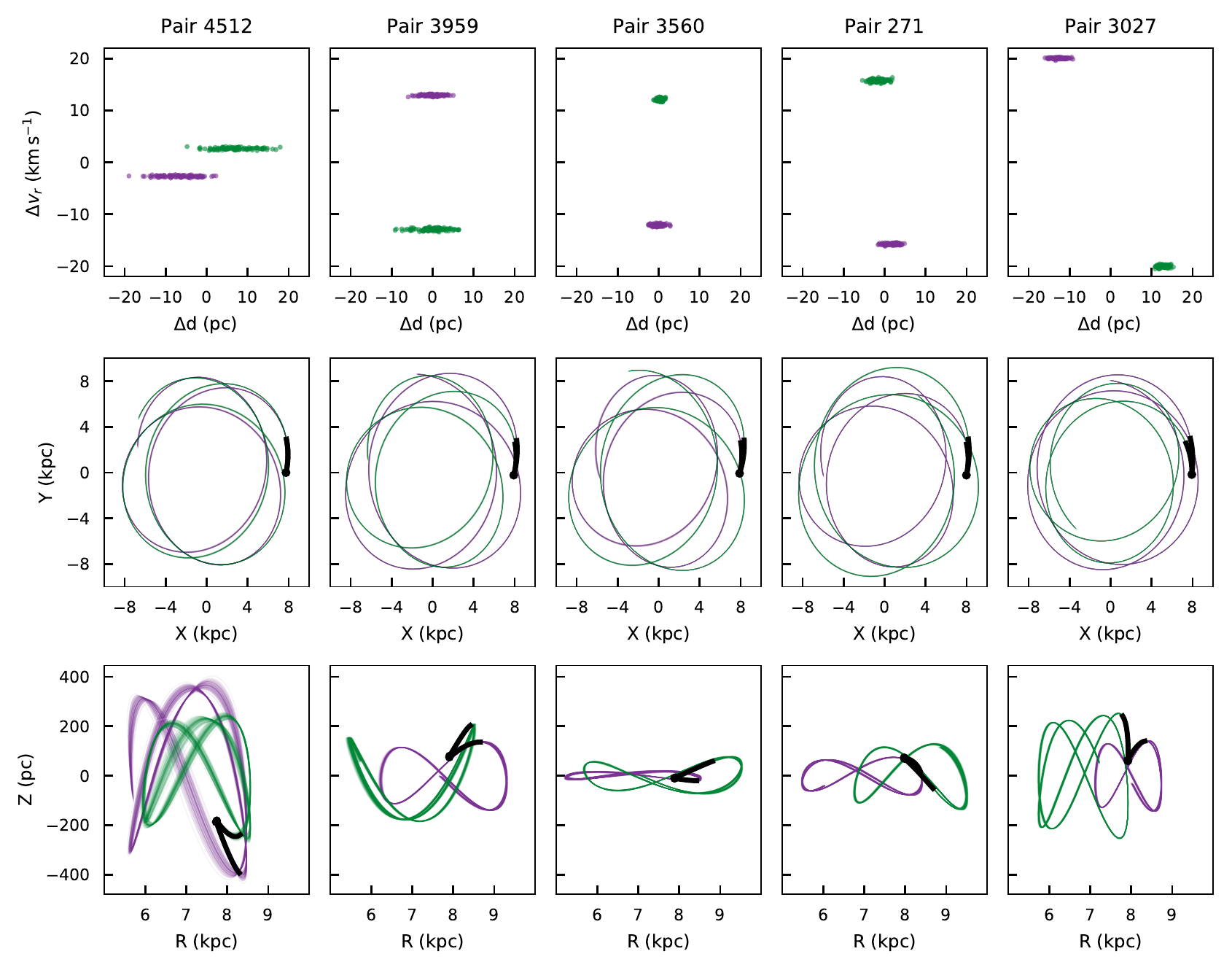}
    \caption{Projections of the orbits of the five pairs with the largest radial velocity difference. For each star, 100 versions of its orbit are shown where the input parameters were drawn from a multivariate normal distribution which took into account the covariances between the parameters and their uncertainties. The top row shows the relative radial velocities of and separations between the stars in each \citetalias{Oh2016} pair; the second row is the projections of their Galactic orbits in the X-Y; and the third row is their R-Z planes. On the orbit plots, a black dot indicates the current median position of the stars and the black lines show the median direction of motion. In all cases, the orbit integrations are consistent with none of these pairs being co-moving about the Galaxy.}
    \label{fig:orbit_pairs_uncertaities_2}
\end{figure*}

\begin{figure*}
    \includegraphics[width=\textwidth]{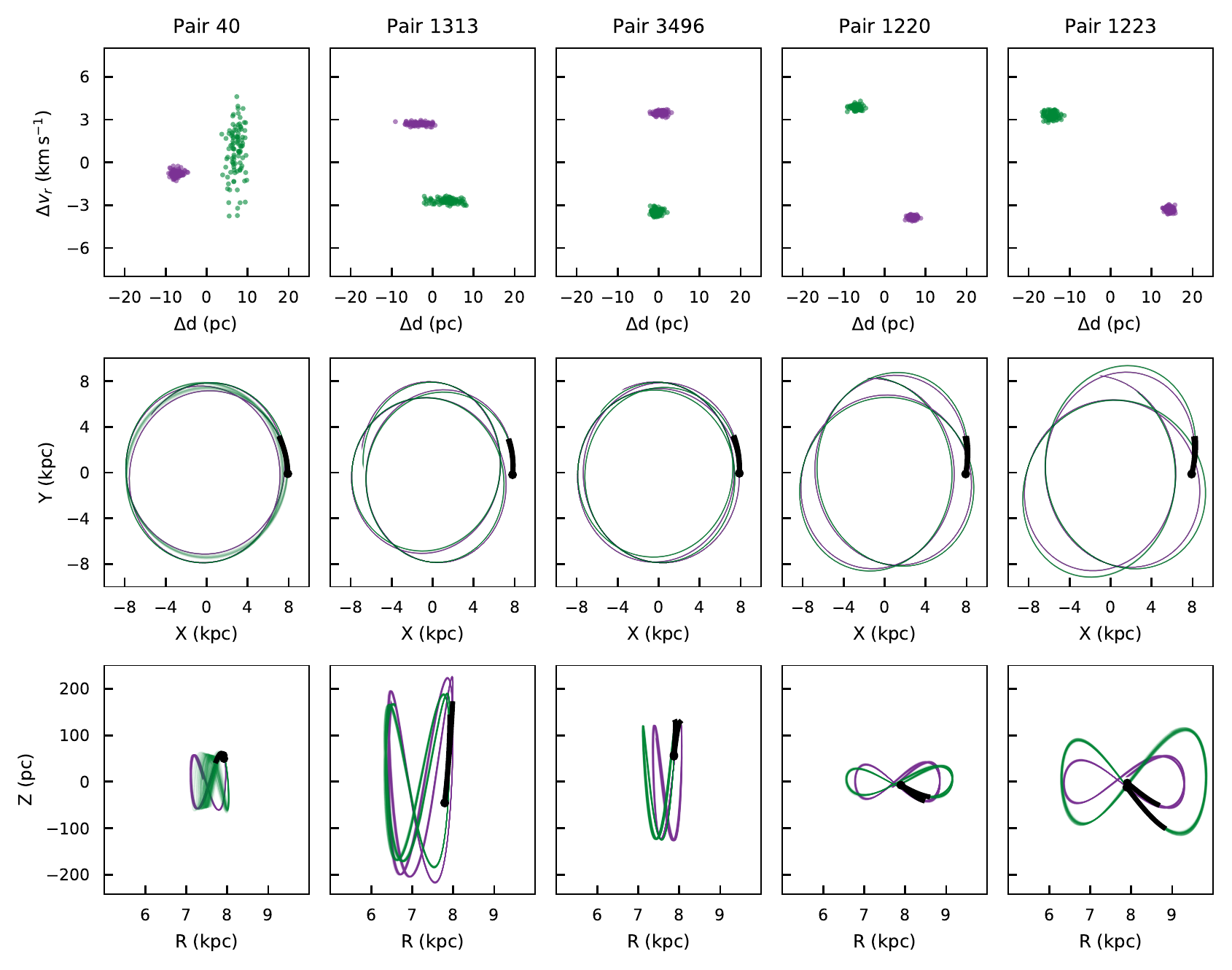}
    \caption{Same as Figure \ref{fig:orbit_pairs_uncertaities_2} but for five pairs of stars with smaller radial velocity differences. One star of Pair 40 has a large uncertainty in its radial velocity. This manifests in the orbit integrations as a larger uncertainty in its perigalaticon than for most other stars considered (Table \ref{table:pair_stellar_params}).} 
    \label{fig:orbit_pairs_uncertaities_1}
\end{figure*}

\begin{figure*}
    \includegraphics[width=\textwidth]{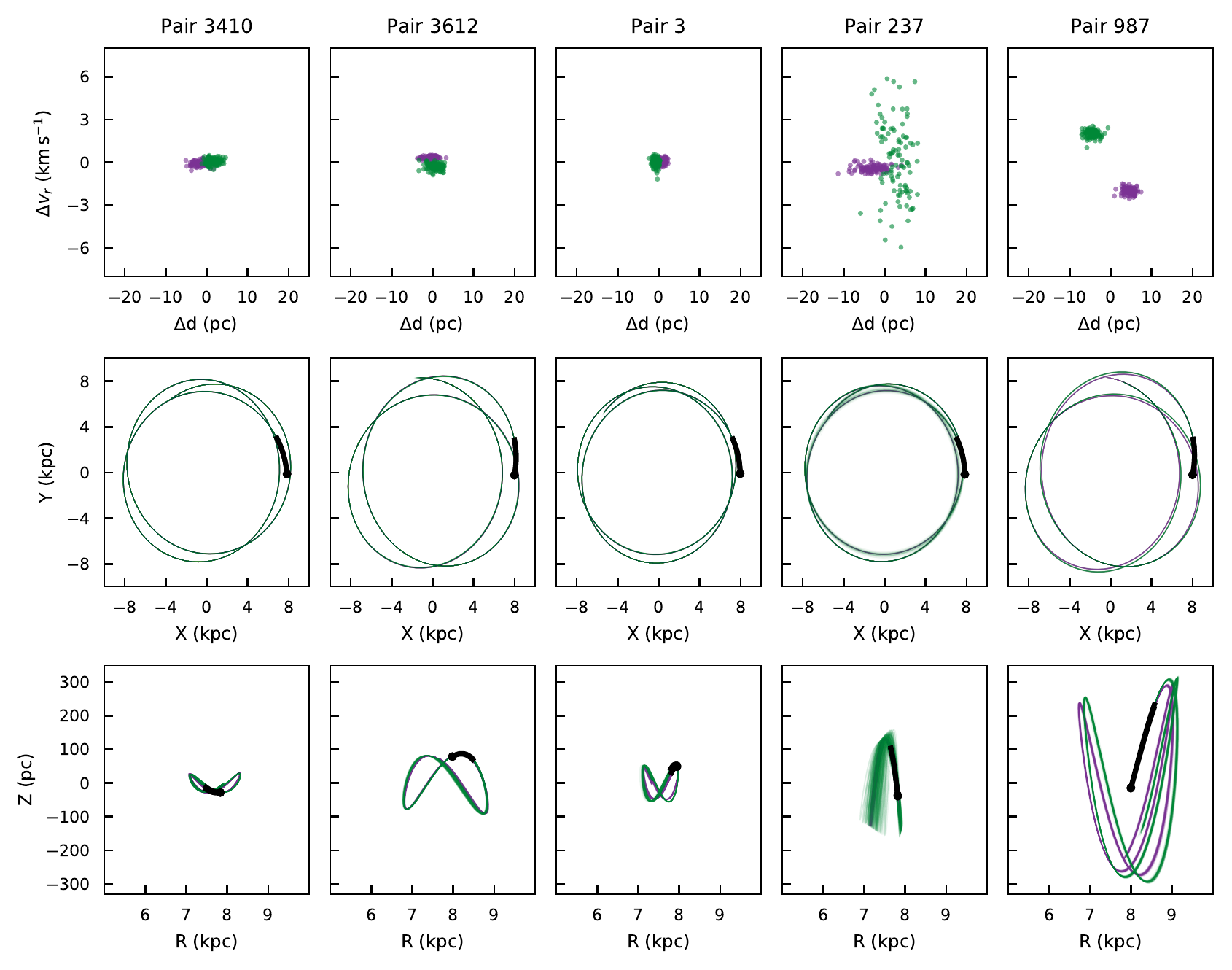}
    \caption{Same as Figure \ref{fig:orbit_pairs_uncertaities_2} but for five pairs of stars with smallest radial velocity differences. One star of Pair 237 has a large RV uncertainty (as for one star of Pair 40; Figure \ref{fig:orbit_pairs_uncertaities_1}). For Pairs 3410, 3612, 3 and 237, these orbit integrations provide good evidence that they are truly co-orbiting the Galaxy.}
    \label{fig:orbit_pairs_uncertaities_0}
\end{figure*}

\begin{figure}
    \includegraphics[width=\columnwidth]{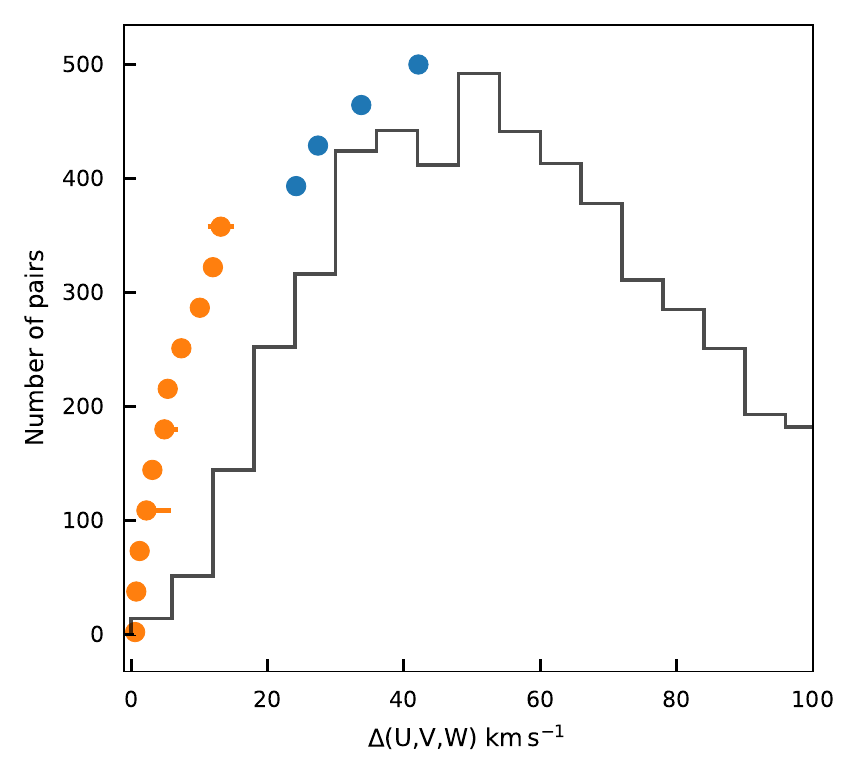}
    \caption{The distribution of $\Delta(U,V,W)$ of the 5755 random GALAH-observed pairs (black histogram) compared to that of the GALAH DR2 pairs in common with \citetalias{Oh2016} (orange and blue dots with errobars; values from Table \ref{table:basic_data}). The background distribution uses pairs that were selected solely to be spatially close ($d<10$~pc), while the \citetalias{Oh2016} pairs are spatially close and have similar proper motions. We find that four of the 15 pairs have $\Delta(U,V,W)$ at the peak of the background distribution. The other 11 all have $\Delta(U,V,W)<14$~km\,s$^{-1}$, which was only the case for 2 per cent (111/5755) of the random pairs.}
    \label{fig:uvw_real_nonreal}
\end{figure}

\begin{table*}
\centering
\caption{Differences in position and velocities of 15 pairs of stars found in GALAH that were identified in \citetalias{Oh2016}. The values are the median, and the 5th and 95 percentiles of 10000 random samples of the 6D information of the stars taking account their uncertainties and covariances in \textit{Gaia} DR2. The pairs are sorted by the $\Delta(\mathrm{U,V,W})$ difference of the pair. Pairs above the line have lower $\Delta(\mathrm{U,V,W})$ and are potentially co-orbiting pairs, while below the line are likely not. All but one has a physical separation greater than 3.1~pc.}
\label{table:basic_data}
\begin{tabular}{rrrrrrrr}
\hline 
Pair ID & ang sep (arcmin) & separation (pc) & $\Delta(v_r)$ (km/s) & $\Delta(U)$ (km/s) & $\Delta(V)$ (km/s) & $\Delta(W)$ (km/s) & $\Delta(\mathrm{U,V,W})$ \\
\hline
3410 & $1$ & $3.1\substack{+3.3 \\ -2.7}$ & $0.1\pm0.4$ & $-0.1\pm0.4$ & $-0.2\pm0.3$ & $-0.5\pm0.2$ & $0.6\pm0.2$ \\
3612 & $3$ & $1.5\substack{+2.8 \\ -1.2}$ & $0.6\pm0.4$ & $-0.1\pm0.8$ & $0.5\pm0.4$ & $-0.3\pm0.2$ & $0.8\pm0.5$ \\
3 & $110$ & $4.2\substack{+0.8 \\ -0.4}$ & $0.1\pm0.6$ & $-0.9\pm0.3$ & $0.1\pm0.5$ & $-0.8\pm0.2$ & $1.3\pm0.4$ \\
237 & $88$ & $8.6\substack{+5.0 \\ -2.1}$ & $0.8\pm4.9$ & $-0.3\pm3.6$ & $-0.8\pm3.1$ & $-0.4\pm1.2$ & $2.2\pm3.6$ \\
987 & $51$ & $9.6\substack{+2.8 \\ -2.7}$ & $3.9\pm0.5$ & $0.1\pm0.9$ & $-2.9\pm0.5$ & $-0.9\pm0.2$ & $3.1\pm0.5$ \\
40 & $185$ & $16.4\substack{+2.2 \\ -2.1}$ & $1.7\pm2.8$ & $-1.0\pm1.8$ & $-4.7\pm2.1$ & $-0.7\pm0.6$ & $4.9\pm2.0$ \\
1313 & $44$ & $7.9\substack{+4.5 \\ -3.7}$ & $5.4\pm0.3$ & $-4.4\pm0.6$ & $1.9\pm0.6$ & $2.3\pm0.2$ & $5.4\pm0.4$ \\
3496 & $69$ & $3.2\substack{+1.1 \\ -0.2}$ & $6.9\pm0.3$ & $-5.3\pm0.3$ & $5.2\pm0.4$ & $0.0\pm0.1$ & $7.4\pm0.3$ \\
1220 & $112$ & $14.8\substack{+1.8 \\ -1.7}$ & $7.7\pm0.3$ & $10.0\pm0.6$ & $0.4\pm0.4$ & $-0.7\pm0.2$ & $10.1\pm0.6$ \\
1223 & $111$ & $29.2\substack{+2.1 \\ -2.0}$ & $6.6\pm0.4$ & $10.9\pm0.8$ & $-3.5\pm0.6$ & $3.7\pm0.3$ & $12.0\pm0.7$ \\
4512 & $46$ & $13.8\substack{+10.6 \\ -8.6}$ & $5.3\pm0.4$ & $-1.4\pm1.2$ & $-6.9\pm1.1$ & $-11.1\pm1.5$ & $13.1\pm2.0$ \\
\hline
3959 & $101$ & $7.8\substack{+2.3 \\ -0.4}$ & $25.8\pm0.4$ & $-11.0\pm1.7$ & $20.9\pm0.8$ & $-5.1\pm0.2$ & $24.2\pm0.5$ \\
3560 & $172$ & $6.9\substack{+0.7 \\ -0.1}$ & $24.2\pm0.4$ & $21.1\pm0.8$ & $-16.3\pm1.1$ & $-6.3\pm0.3$ & $27.4\pm0.5$ \\
271 & $86$ & $6.9\substack{+2.3 \\ -1.0}$ & $31.5\pm0.4$ & $11.1\pm0.9$ & $-31.4\pm0.5$ & $5.6\pm0.3$ & $33.8\pm0.5$ \\
3027 & $48$ & $25.8\substack{+3.1 \\ -3.0}$ & $40.1\pm0.4$ & $-28.2\pm0.6$ & $30.1\pm0.5$ & $-8.4\pm0.1$ & $42.1\pm0.4$ \\
\hline
\end{tabular}
\end{table*}

Our first step in evaluating whether the possible \citetalias{Oh2016} pairs are truly co-moving was to integrate their orbits around the Galaxy. For each star the covariance matrix was constructed from the reported errors and covariances in \textit{Gaia} DR2, and then 100 samples were drawn using \texttt{numpy.random.multivariate\_normal} to give the RA, Dec, the inverse parallax, proper motions in RA and Dec, and the radial velocity $(\alpha,\delta,r_\textrm{\sun},\mu_\alpha\cos\delta,\mu_\delta,v_r)$. It is important to consider the uncertainties of these values as it is not intuitive how a large uncertainty in one parameter will impact the orbit, especially as we are taking the projected velocities on the sky. An orbit was computed for each sample using \textsc{galpy} \citep[\url{http://github.com/jobovy/galpy};][version 1.3]{Bovy2015} with the recommended Milky-Way-like \texttt{MWPotential2014} potential, and the Solar motion defined by \citet{Schonrich2010}. The orbits were integrated forward in time for 500~Myr with 0.5~Myr resolution. Note that these orbital integrations do not take into account the mutual gravitational attraction of the pairs.

We show projections of 100 orbit samples for each star of each pair in Figures \ref{fig:orbit_pairs_uncertaities_2}, \ref{fig:orbit_pairs_uncertaities_1}, \ref{fig:orbit_pairs_uncertaities_0}. Each star in each pair is plotted in a different colour. For each star we show the radial velocities of the stars in the pair relative to the mean versus the distances of the stars in the group relative to the mean; the integrated orbits projected into the Galactic X-Y plane; and the orbits in the R-Z plane. In the orbit panels, the current position of each star is shown with a black dot, and the first 15 Myr of its orbit is shown with a black line, to indicate the direction of motion. In Table \ref{table:pair_stellar_params} we give the eccentricity, maximum vertical height, perigalacticon, and apogalacticon for these orbits. In most cases the uncertainties and covariances of the input parameters do not manifest as very uncertain orbits. The counter-example is Pair 4512 which has a large uncertainty in the stellar distances, which causes a range of possible future orbits. It is important to consider the uncertainties and covariances of the input parameters as it is not intuitive how a large uncertainty in a given parameter will impact the orbit.

The thin disc is dynamically fairly cold; that is, the velocity dispersion in the (U,V,W) velocity space is not very large, and the orbits of individual stars tend to have low eccentricity and be confined to the plane of the disc \citep[e.g.,][]{Edvardsson1993, Pasetto2012}. This introduces the possibility that stars might be close to each other and co-orbiting without having formed together. To better understand how random, but spatially close, pairs of stars look in $(\mathrm{U,V,W})$ space, we carry out a simple experiment within the GALAH dataset (Table \ref{table:basic_data}). We identified 5755 pairs of stars in GALAH within 600~pc of the Sun, for which each star has only one other star within 10~pc (i.e., not in clusters or large associations), and calculate $\Delta(\mathrm{U,V,W})$, the Cartesian distance between the velocities of the two stars. Figure \ref{fig:uvw_real_nonreal} shows the distributions of $\Delta(\mathrm{U,V,W})$ with the values for the co-orbiting stars overplotted. In Table \ref{table:basic_data} we give the spatial and observed kinematic differences between each of the pairs.

For the random pairs, the peak of the $\Delta(\mathrm{U,V,W})$ is about $40~\mathrm{km}\,\mathrm{s}^{-1}$. The pairs of stars we consider from \citetalias{Oh2016} tend to have lower $\Delta(\mathrm{U,V,W})$ than the bulk of the random pairs. This is not surprising; our selection of the random pairs only required them to be spatially close, while the \citetalias{Oh2016} pairs were both spatially close and similar in proper motion. Four of the 15 pairs have large $\Delta(\mathrm{U,V,W})$, at values like the peak of the distribution of $\Delta(\mathrm{U,V,W})$ for the random pairs. These four pairs (271, 3027, 3560 and 3959, shown in Figure \ref{fig:orbit_pairs_uncertaities_2}) have the largest radial velocity differences, with  $|\Delta\mathrm{RV}|>24$~km\,s$^{-1}$. Their integrated orbits show that within these pairs, the stars have widely different inclinations or eccentricities. We, therefore, conclude that the apparent association in proper motion of the stars in these four groups is coincidental.

For the other 11 pairs (Figures \ref{fig:orbit_pairs_uncertaities_2}, \ref{fig:orbit_pairs_uncertaities_1}, \ref{fig:orbit_pairs_uncertaities_0}), we find that some do appear very likely to be co-moving pairs. The difference in radial velocity for all of these pairs is $<7.7~\mathrm{km}\,\mathrm{s}^{-1}$, and in four cases is $<1.0~\mathrm{km}\,\mathrm{s}^{-1}$. These four pairs (Figure \ref{fig:orbit_pairs_uncertaities_0}) we consider to be the most likely to be co-moving. The orbits of the stars in the pairs with the largest relative radial velocity do diverge in the R-Z plane (Figure \ref{fig:orbit_pairs_uncertaities_1}), especially pairs 4512 and 1223, so although they have similar orbits now they may not remain associated in the future. \textcolor{red}{\citet{Andrews2018} predicts, and \citet{Andrews2018b} demonstrates, that the majority of apparent co-moving pairs in TGAS with separations larger than 0.2 pc are not truly co-orbiting. All of the potentially co-moving pairs in this study have separations larger than that limit. While we find that a minority of the 15 pairs we consider are not co-orbiting, this is a small enough data set that we cannot strongly support or contradict the 0.2 pc limit described by \citet{Andrews2018}}. In all cases, more detailed orbit integrations (i.e., more realistic potential, considering the gravitational interaction between the stars) would provide a more conclusive answer to whether these stars will continue to orbit together. While we are considering only a small subset of the \citetalias{Oh2016} sample, it is clear that full 6D velocity confirmation is necessary before drawing larger conclusions about cluster dissolution or disc substructure from the reported co-moving pairs.

\subsection{Similarity in stellar parameters, photometry and spectra}\label{sec:cmd}

\begin{figure}
    \includegraphics[width=\columnwidth]{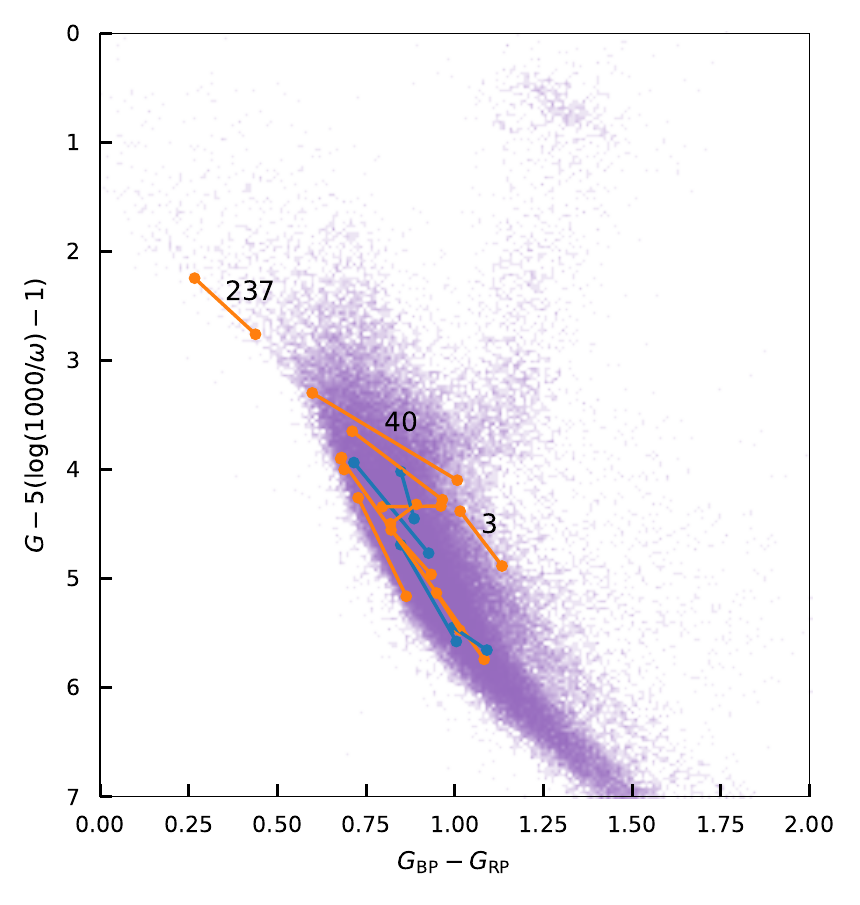}
    \caption{Colour-magnitude diagram of all stars observed by GALAH within about 600~pc of the Sun. Also plotted are the fifteen stellar pairs from \citetalias{Oh2016} that were serendipitously observed as part of the GALAH survey. Those pairs that GALAH data show are kinematically similar are shown in orange, and those that are dissimilar are shown in blue. Each pair is connected by a line. The three numbered pairs are discussed in Section \ref{sec:cmd}.}
    \label{fig:pairs_cmd}
\end{figure}

\begin{figure*}
    \includegraphics[width=\textwidth]{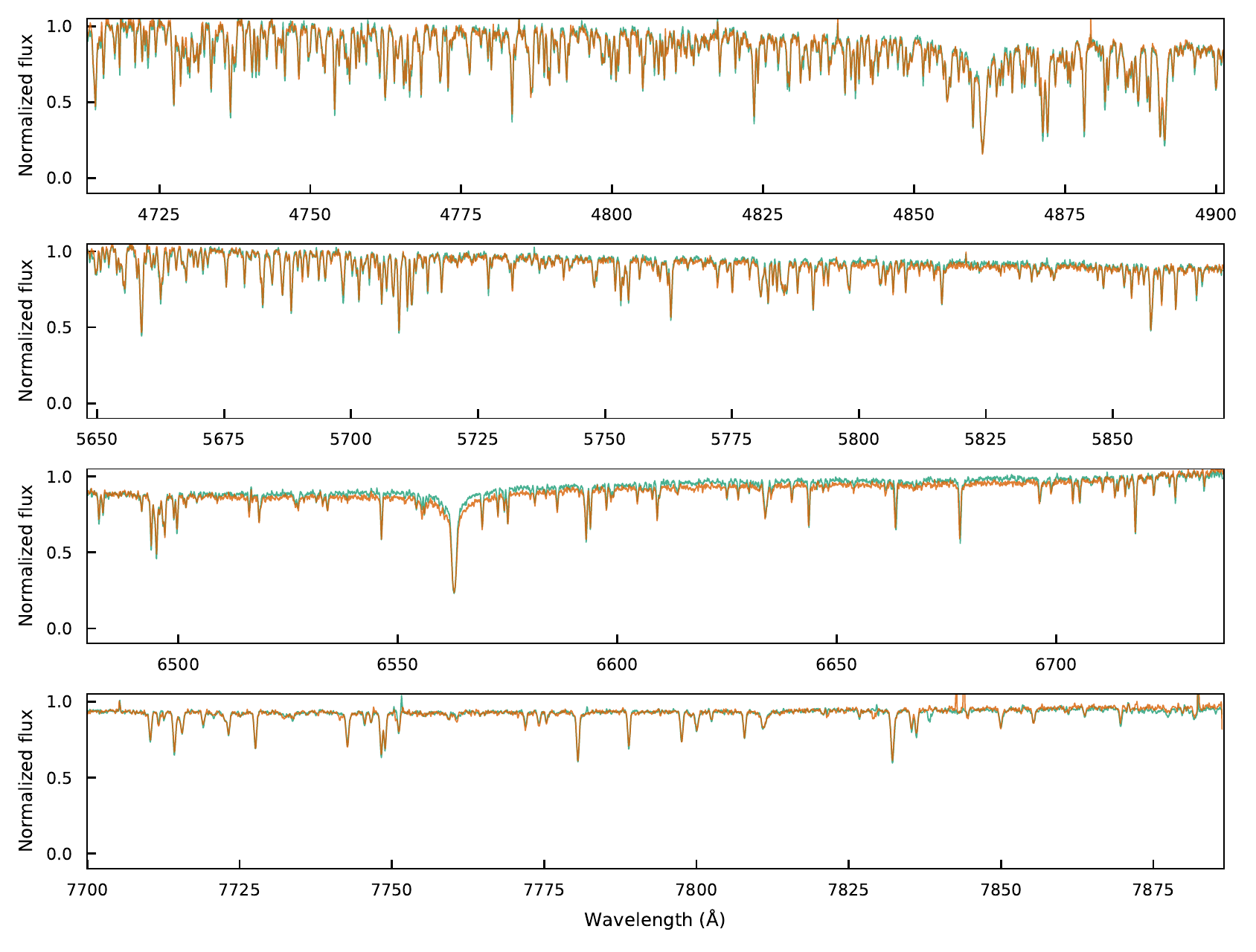}
    \caption{Spectra of the two stars of Pair 3612 in the four HERMES bandpasses. The spectra are so similar that it is difficult to distinguish them. In Figure \ref{fig:spec_pair_40_zooms}, we show small cutouts of some of the spectral lines used in the abundance determination.}
    \label{fig:spec_pair_example}
\end{figure*}

\begin{figure}
    \includegraphics[width=\columnwidth]{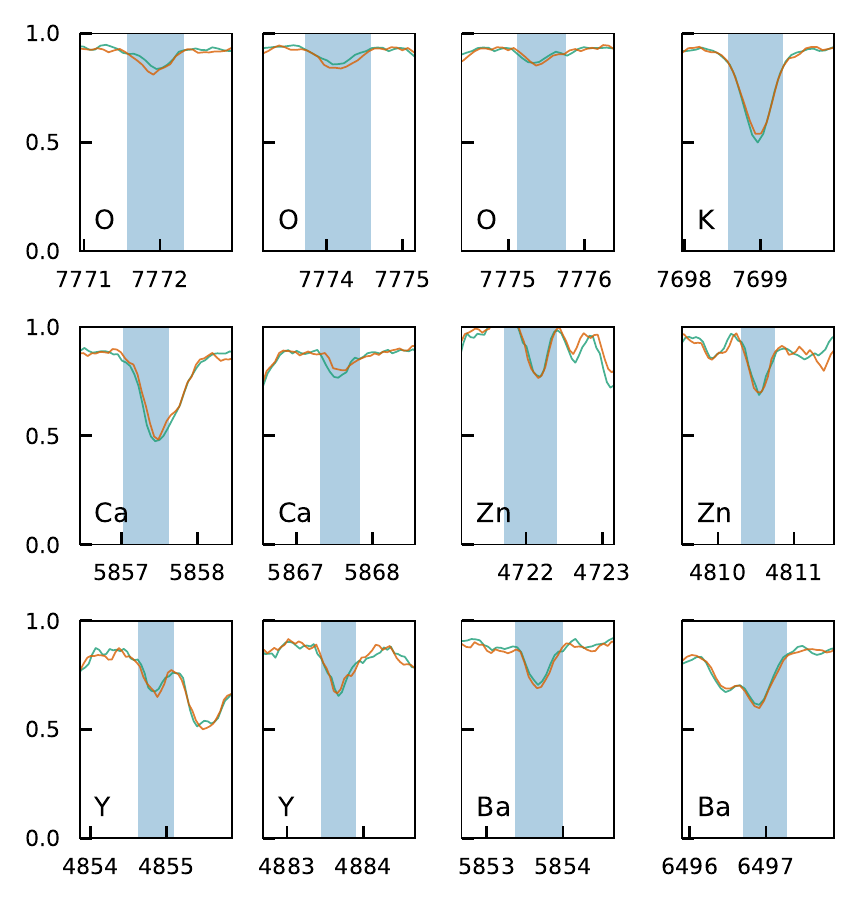}
    \caption{Assorted cutouts from Figure \ref{fig:spec_pair_example} of some of the spectral regions used for the abundance determination, highlighting how very similar the spectra of Pair 3612 are. The blue shaded region in each panel is the mask used by \textit{The Cannon} for that line.}
    \label{fig:spec_pair_40_zooms}
\end{figure}

In several of the co-orbiting pairs, the stars have similar magnitudes and colours (which are given in Table \ref{table:pair_stellar_params}). Since they were selected to have similar distances from the Sun, we expect that their stellar parameters should also be close. Figure \ref{fig:pairs_cmd} shows the (absolute $G$, $G_\mathrm{BP}-G_\mathrm{RP}$) colour-magnitude diagram for all GALAH stars within 600 pc of the Sun. The low $\Delta(\mathrm{U,V,W})$ pairs are highlighted as larger orange dots and the four kinematically dissimilar pairs as larger blue dots, and each pair is connected with a line. Unsurprisingly for apparently bright stars located relatively nearby, the stars in the pairs tend to be on the main sequence. There is one pair that contains a potential subgiant (Pair 40). Rather than stellar twins, almost all of the pairs consist of a brighter, hotter star with a fainter, cooler ``companion''. This is consistent with the pairs of stars being the same age and differing slightly in mass, but it is not conclusive, since a) the majority of stars within 600~pc of the Sun are main-sequence stars, so randomly assigned pairs would tend to behave in this same way, and b) the kinematically dissimilar pairs show similar arrangements.

There are three pairs for which we can evaluate ages slightly more precisely: groups 237, 3, and 40. Group 237 is the most luminous main-sequence pair in our data set, and given the shorter main sequence lifetimes for higher-mass stars, the stars in this group must be closer in age than the other pairs. Comparison with a solar metallicity MIST isochrone \citep{Dotter2016,Choi2016,Paxton2011,Paxton2013,Paxton2015} returns a maximum age for the brighter star in group 237 of about 800~Myr.

Meanwhile, the stars from Group 3 are distinctly redder than most of the other pairs for their luminosity. These stars are members of the Sco-Cen OB association and their main sequence is brighter at redder colours than the bulk of the stars in our sample because of their young age. Isochrone fits yields an age of about 10~Myr for this pair, which is consistent with age estimates for the association \citep[e.g.,][]{Pecaut2012}. It is likely that they are pre-main sequence stars, so it is unsurprising that \textit{The Cannon}, which was not trained for that evolutionary stage, has not returned reliable parameters for either star.

Group 40 has one main-sequence member and one apparent subgiant member, but it was not possible to fit a single isochrone to both stars. The subgiant star is faint enough that it must have lower mass and higher age than the main sequence star, even though they are likely co-orbiting. Unfortunately, we only have reliable label flags for the main-sequence star, so cannot confirm this with stellar parameters and abundances.

In some cases, the stars are similar enough that a direct comparison of the spectra is a sufficient demonstration of highly similar abundance patterns \citep[similar to what was shown in][]{Bovy2016}. The stellar parameters derived for the two stars in group 3612 are practically identical. Their full spectra are shown in Figure \ref{fig:spec_pair_example}, and small cut-outs of spectral regions used for the abundance determination in Figure \ref{fig:spec_pair_40_zooms}. The spectra of the two stars are remarkably alike, which, together with the matching parameters, indicates that the two stars must have quite similar abundance patterns. Deriving stellar parameters and abundances for two stars that are clearly so similar in their observational properties is a good basic verification of the GALAH analysis process. This is explored in the next section.

\section{Abundance behaviour in co-orbiting pairs}\label{sec:abundances}

\begin{figure*}
    \includegraphics[width=0.99\textwidth]{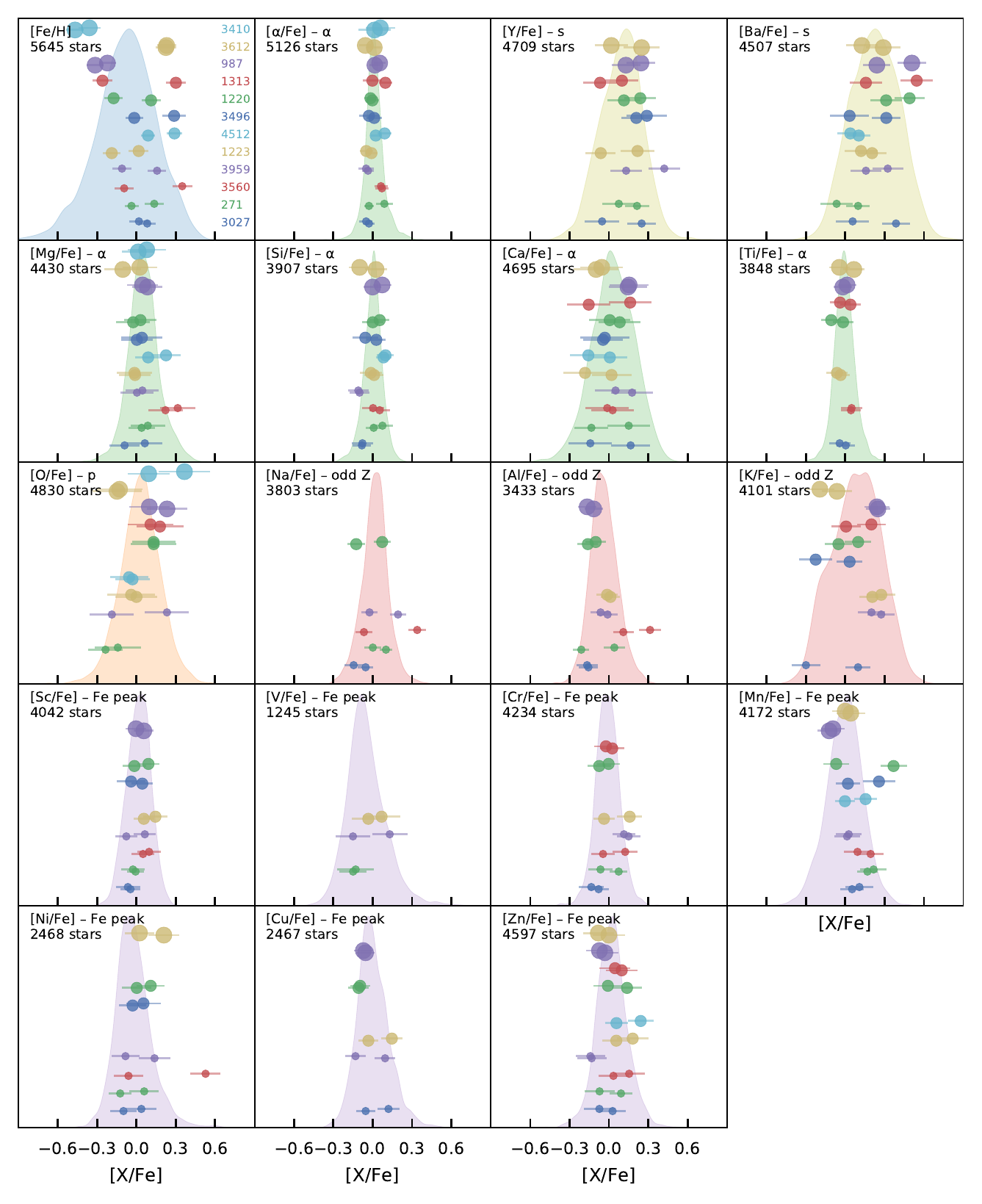}
    \caption{Abundance patterns for the pairs in common between GALAH and \citetalias{Oh2016}. Large circles are used for the low $\Delta(\mathrm{U,V,W})$ pairs (i.e., $\Delta(\mathrm{U,V,W}))<20$~km\,s$^{-1}$) and small symbols for the non-co-moving pairs. The colouring of the pairs is simply to aid the eye. The vertical ordering is by the difference in $\Delta(\mathrm{U,V,W})$ velocity of the pair, from smallest-to-largest top-to-bottom, and is the same in all panels. Some pairs are missing from some panels because that particular element was flagged as unreliable in one or both of the stars. The background distribution is for all GALAH stars within about 600~pc of the Sun and the number of stars that make up the distribution is given in each panel.}
    \label{fig:abund_hist_real}
\end{figure*}

\begin{figure*}
    \includegraphics[width=\textwidth]{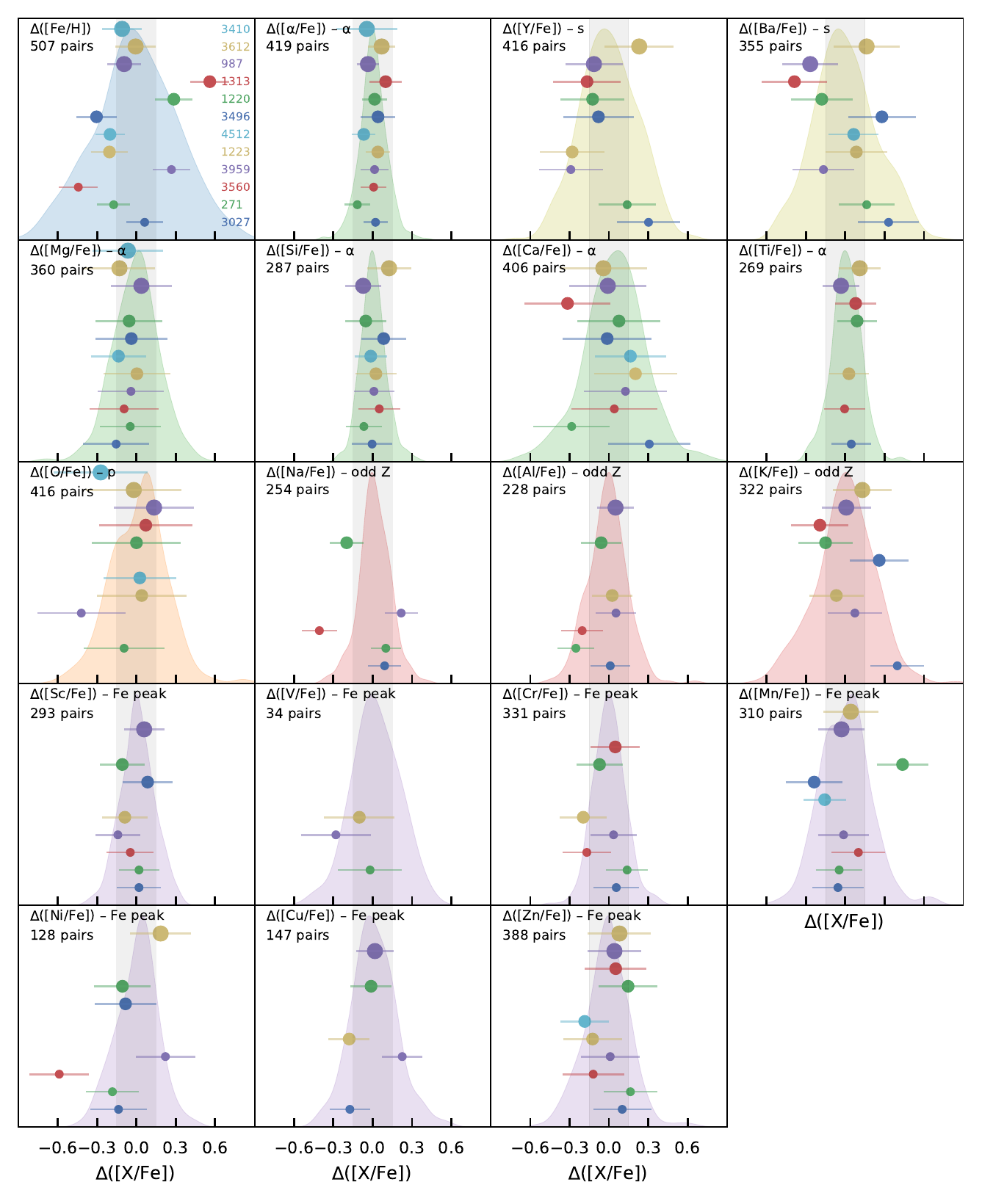}
    \caption{Similar to Figure \ref{fig:abund_hist_real}, but showing the difference in the abundances for each pair. The background distributions are the abundance differences of the 5755 random, spatially close pairs selected from GALAH, with each panel giving the number of useful pairs for that element. Errorbars are the sum of the errors of the pair. The shaded vertical region show $\Delta\mathrm{[X/Fe]}\pm0.15$~dex. Only the top three pairs, and the bottom pair show similar metallicities.}
    \label{fig:abund_diff}
\end{figure*}

The chemical homogeneity of stars that form in the same environment at the same time is an axiom of star formation; considering gas clouds that will collapse to form stars, \citet{Feng2014} showed that turbulent mixing is highly effective at homogenising the composition of higher-mass clouds. Lower-mass clouds and star-forming clouds in regions with a lower star formation efficiency may potentially be less well-mixed, but they are not as well studied. However, when abundances of stars in open clusters are measured at very high precision, there are clear inhomogeneities. Using high-precision differential abundances measured from very high-quality spectra of stars in the Hyades, \citet{Liu2016} found that there are star-to-star abundance variations on the order of 0.02 dex and that most of the elemental abundances in each star are correlated to each other.

We expect, therefore, to see a baseline of homogeneity for stars that formed together, to within the precision of the GALAH abundances ($\approx 0.1$ dex) and we do not expect this for stars co-orbiting due to dynamical effects (like stars in the Hercules stream, which are co-orbiting as a result of resonance with the Galactic bar). However, it is important to remember that there are reasons that co-natal, co-eval stars might have mismatched abundances, e.g., atomic diffusion \citep[e.g.,][]{Dotter2017}, planet formation \citep[e.g.,][]{Melendez2009}, and binary mass transfer \citep[e.g.,][]{Hansen2015}, all of which affect a particular set of elemental abundances during particular evolutionary phases. As one example of mismatched but potentially co-natal stars, \citet{Oh2018} used archival Keck/HIRES spectra for pair 1199 from \citetalias{Oh2016}, and found that the two stars have very similar 3D velocities, and are consistent with having similar ages. However, they found that one star is enhanced by 0.2 dex in refractory elements and by 0.05--0.10 dex in volatile elements relative to the other star. They interpret this pair of stars as having formed together, with the relatively enhanced star having later accreted rocky material, presumably following the formation of a planetary system.

In 12 of the 15 pairs, both stars have $\mathrm{\texttt{flag\_cannon}}=0$, meaning that their stellar labels have no evidence for being untrustworthy (unfortunately the missing three are some of most kinematically similar pairs). We plot their metallicity, [$\alpha$/Fe]\footnote{In GALAH, [$\alpha$/Fe] is the error-weighted combination of Mg, Si, Ca, Ti abundances. See \citet{Buder2018}.}, and elemental abundances on Figures \ref{fig:abund_hist_real} and \ref{fig:abund_diff}, with one panel per element and the same horizontal scale in all panels. Figure \ref{fig:abund_hist_real} plots the [X/Fe] values for each star of each pair, while Figure \ref{fig:abund_diff} plots the abundance differences --- $\Delta(\mathrm{[X/Fe]}$ --- of each pair. Large circles are used for small $\Delta(\mathrm{U,V,W})$ pairs and small symbols for the non-co-moving pairs. The colour of each pair are merely to aid the eye and they have the same vertical arrangement in each panel, sorted by  $\Delta(\mathrm{U,V,W})$. The group numbers from \citetalias{Oh2016} are listed to the right of each group in the panel showing [Fe/H]. In some cases, individual element abundances in one or both stars of a pair were flagged as unreliable, so some pairs are missing from some panels. 

The elements are arranged by their dominant nucleosynthetic groups, represented by the colour of the background histogram, with the group name (``p'' for proton capture, ``odd Z'' for light odd-Z elements, ``$\alpha$'' for alpha elements, ``s'' for the slow neutron capture process, ``r'' for the rapid neutron capture process and ``Fe peak'' for the iron-peak elements) next to the name of the element. In Figure \ref{fig:abund_hist_real} the distribution of each element's abundance for all GALAH targets within about 600~pc of the Sun is shown as a smoothed histogram in the background of each panel. In Figure \ref{fig:abund_diff} we show the distribution of differences for the 5755 random, spatially close pairs drawn from GALAH within 600~pc of the Sun. The number of stars or pairs in the background histogram is given in the upper left of each panel, since not all elements are measured in all stars.

In Figure \ref{fig:abund_hist_real}, the background distributions of $\alpha$ and iron-peak abundances are quite narrow, indicating that type Ia supernov\ae\ played a strong role in enriching the gas that formed these stars. This is not a surprise since the stars are all within about 600~pc of the Sun, and (based on our orbital integrations) typically associated with the thin disc. The slightly broader ranges in the proton capture and s-process elements suggests that feedback from low- and intermediate-mass stars also contributed to the chemical evolution of the material from which the stars formed. This is also consistent with these stars being relatively young thin disc stars since their late formation allows time for (as an example) AGB stars to produce and eject significant amounts of CNO-cycle and s-process elements. These broader distributions may also reflect their larger abundance uncertainties, as there are typically fewer spectral lines being used for these elements \citep{Buder2018}.

Looking at our pairs of stars in the \feh\ panel of Figure \ref{fig:abund_hist_real}, we find that they are evenly distributed across the background distribution. In most cases, for both the kinematically similar and dissimilar pairs, there are large metallicity differences between the stars in the pair: typically $\Delta\feh>0.15$ (Figure \ref{fig:abund_diff}). This is a much larger difference than would ordinarily be explained by inhomogeneities in the pre-stellar nebula, planet formation \citep[e.g.,][]{Melendez2017}, or differential effects of diffusion on main-sequence stars with different masses \citep[e.g.,][]{Michaud2004}. There are dynamical processes that could result in non-co-natal stars occupying matching orbits, stellar captures or partner swapping in binaries/triple systems, that might explain the pairs with very similar orbits but differing metallicity.

Pairs 3410, 3612 and 987 are the only kinematically similar pairs that have $\Delta\feh<0.11$. Looking at the other abundance difference panels of Figure \ref{fig:abund_diff}, these are the only pairs that are consistently similar within the uncertainties. In general, the abundance difference between the stars in each pair is smaller than the spread of the background distribution, reinforcing the claim from the orbital properties that the stars in each pair are intrinsically associated. Considering Pair 3612 which we highlighted in Figure \ref{fig:spec_pair_example}, the results shown in Figure \ref{fig:abund_diff} find that only in [Y/Fe] does the difference fall outside of the $\pm0.15$~dex range.

Pair 3027 has the largest velocity difference, but its stars have similar \feh. This is likely a coincidence as both stars in this pair sit at the most common value of the underlying metallicity distribution --- slightly super-solar. It has abundance differences for several of the elements determined by GALAH. All of the other kinematically dissimilar pairs have large metallicity differences, confirming the results from the radial velocities that these are neither co-orbiting, nor co-natal pairs.

We warn the reader against over-interpreting these abundance data. The background distributions of the alpha and iron-peak elements are narrow enough that it can be difficult for those abundances to be significantly different, even in randomly chosen pairs. They are not that informative for chemical tagging to distinguish between different co-orbiting pairs, then, but since the background distributions are so narrow, they would be very effective for distinguishing the thin disc from any thick disc \citep[e.g.,][]{Bensby2014} or halo \citep[e.g.,][]{Roederer2010} stars in the data set.

\section{Discussion}\label{sec:discussion}
In this work, we have presented a fundamental step in the development of detailed chemical tagging and explored our ability to distinguish true co-orbiting stars from those that are coincidentally kinematically close using spectroscopic stellar parameters and abundances. We reiterate that the main driver of the GALAH survey is chemical tagging solely in abundance space. As discussed in Section \ref{sec:introduction} we aim to search for co-eval and co-natal stars that have lost their coherence in kinematic space due to being dispersed throughout the Galaxy. In this work, however, stars were initially selected from their phase space information, and then the GALAH spectra and abundances were used to confirm a (or refute) similar common origin for the stars.

The small size of the data set considered in this study allows the star-by-star analysis we have chosen to use, but full chemical tagging in the Galactic disc, like many other goals of Galactic archaeology, will require a fairly high level of automation based on well-justified metrics and statistics. There have been several methods already proposed and tested, including the Manhattan distance metric \citep{Mitschang2013}, t-SNE dimensionality reduction \citep[][]{Kos2017a,Anders2018}, principal component analysis \citep{Blanco-Cuaresma2015}, k-means clustering \citep{Hogg2016}, deriving the chemical dimensionality from the spectra in a semi-model independent method \citep{Price-Jones2017}, and unsupervised clustering \citep[including a minimum spanning tree;][]{Boesso2018}.

The strengths of these various methods will make them more or less suited for particular chemical tagging problems. For example, k-means or extreme deconvolution \citep{Bovy2010} will assign all stars in a data set to a given number of groups, and are therefore ideal for disentangling multiple known populations mixed together, while t-SNE and a DBSCAN clustering \citep{Traven2017} will identify all groups above a certain density threshold. PCA is very effective at identifying baseline trends in abundance space but can be derailed by outliers.

The challenges in applying these methods come both from physics and from data analysis. Galactic chemical evolution is extremely complex, with many sources of enrichment contributing differently, but that does not necessarily guarantee that each star formation site over its history has had a unique abundance pattern \citep[e.g.,][]{Ness2018}. A low level of intrinsic abundance scatter has been seen in Galactic open clusters \citep{Liu2016}, and the level of that scatter is theoretically expected to be a function of cluster mass \citep{Feng2014}. Furthermore, even if stars begin with perfectly identical abundance patterns, atomic diffusion will deplete some of those elements in main-sequence stars, with larger effects in higher-mass stars \citep[e.g.,][]{Gao2018}. First dredge-up then restores the abundances to their pre-diffusion values, creating a variable abundance offset between co-natal dwarfs and giants. 

Spectroscopic analysis introduces a range of uncertainties to chemical tagging. Each elemental abundance value has an error bar that depends on how precisely the observed absorption features can be fit by synthetic spectra, which is affected by the signal to noise ratio and dispersion of the spectrum. The model atmospheres from which synthetic spectra are calculated are not perfect captures of the real physical properties of stellar atmospheres, with non-local thermodynamic equilibrium and 3D atmospheric effects making significant differences in the calculated abundances of some elements \citep[e.g.,][]{Lind2017}. There can also be systematic differences in the abundances determined for dwarf and giant stars \citep[e.g.,][]{Korn2010}.

A complete chemical tagging method will need to account for these factors, interpreting the abundance data using stellar evolutionary models to account for factors like diffusion and abundance evolution, building on a probabilistic model of Galactic chemical evolution and the intrinsic abundance scatter among co-natal stars, and accounting appropriately for systematic and random uncertainties in the measured abundances. Future GALAH data releases will use the isochrone matching code \textsc{elli}, which now yields an estimate of the initial composition \citep{Lin2018}.

This method will also need to consider some fairly fundamental questions: which elements need to be considered? How should their importance be weighted? Should multiple elements from each nucleosynthetic group be counted together or separately? Should we attempt to find similar stars in an [X/H] chemical space or an [X/Fe] chemical space? Just as different chemical tagging problems are better addressed by extreme deconvolution or unsupervised clustering or principal component analysis, they may also be more effectively answered with different baseline choices about which measurements are the most informative.

It is interesting to consider the dynamical history of co-natal stars that are spatially close and on very similar orbits at the present day. Studies of young stellar associations \citep[e.g.,][]{Wright2016} often find diverging velocities, indicating that they will not remain gravitationally bound. The age distribution of open clusters \citep{Friel1995} is highly skewed toward young clusters, indicating that the typical dissolution time for star formation events with masses around $10^{3}~M_{\odot}$ is less than 500~Myr. Lower-mass star formation sites will have a shallower gravitational potential, lowering the escape velocity and leading to a shorter dissolution time. Even if stars that formed together have small relative velocities, scattering interactions with giant molecular clouds or other stars, radial migration, and resonances can all dramatically alter their orbits.

Each star's probability of undergoing some kind of interaction increases with time, such that pairs or groups of co-natal stars that start out on similar orbits will be consistently disrupted over time as one or more stars undergoes some kind of scattering or interaction. For co-natal binary stars, these interactions may result in the exchange of one member of the binary for a field star, potentially producing some of the co-moving pairs that have mismatched abundances. Without a dramatic scattering event, but with even a slight difference in L$_{z}$, unbound but co-orbiting stars will drift apart slowly \citep{Jiang2010}. As a result, we expect that truly co-natal co-moving stars will tend to have completed fewer orbits of the Galaxy than pairs that are coincidentally co-moving or randomly selected spatially close pairs. We can also infer that co-natal, co-moving stars are likely to still be near their initial orbits since interactions that would change their orbits would potentially also disrupt the coherence of their orbits.

The study of \citet{Price-Whelan2017} describes follow-up spectroscopy for 311 potential co-moving pairs from \citetalias{Oh2016}, and finds that 40 per cent of those pairs have highly similar 3D velocities (albeit with radial velocity errors on the order of 5 km~s$^{-1}$). For the 15 pairs of stars considered in this data set, we find that 60 per cent have similar 3D velocities (though with a much smaller sample of stars). However, it is not entirely clear from our orbit integrations that they are truly co-orbiting, or will continue to do so in the future. For example, the stars in group 1313, which have a $\Delta(v_r)$ of $5.4 \pm 0.3$~km\,s$^{-1}$, have essentially the same eccentricity and orbital period, but one star is slightly more bound, with a lower total energy and angular momentum and a $Z_\mathrm{max}$ that is 15 per cent lower than the other star.

There is only one pair of stars in common between our work and \citet{Price-Whelan2017}: \citetalias{Oh2016} Group 3. It is part of the Sco-Cen young stellar association. GALAH DR2 radial velocities are very similar to the RAVE radial velocities adopted for these stars by \citet{Price-Whelan2017}, and since the stars are part of a known association, we agree with their kinematics-based assessment. However, considering stellar parameters and abundances, overall we tend to be more skeptical than \citet{Price-Whelan2017} about the likelihood that the pairs of stars are truly co-natal. Pair 40, for example, cannot originate in the same star formation event because one star must be distinctly older than the other based on their colour-magnitude positions. 

Of the eight co-orbiting pairs for which we have stellar parameters and abundances, three have \feh\ differences of less than $1.5\sigma$, but the other five have \feh\ differences of up to $8\sigma$, or 0.56~dex, and their other abundance differences (as seen in Fig. \ref{fig:abund_diff}) can also be quite large. The abundance differences in these five pairs do not correlate with the elements' condensation temperature in the way that has been interpreted \citep[by, e.g.,][]{Melendez2009} as a sign of rocky planet formation. For these five pairs, although their spatial locations and kinematics are presently quite similar, we conclude that they are not likely to be co-natal. 

Looking ahead, the combination of spectroscopic abundance data from GALAH and other ongoing surveys with \textit{Gaia} kinematics and distances will be extremely powerful. Even in this small sample of 15 reported co-moving groups, we have found three examples of pairs that are consistent with them being from unique formation sites. The large number of abundances available via the GALAH survey allows us to be much more confident with this chemical tagging result than if we had just radial velocities, metallicities and perhaps an $\alpha$-abundance. Testing the assumption that each birth cluster will have a unique set of abundances will be very important. In this data set, comparing Group 3612 with Group 987, 3612 is more metal-rich but relatively depleted in Mg and O, while 987 is more metal-poor with a solar Mg abundance and a small enhancement in O. It is this type of unique abundance profile that we hope to exploit in chemical tagging.

Stars in open and globular clusters are a natural dataset for investigating the prospects for chemical tagging and chemical homogeneity (i.e., it would be hoped that we can chemically tag 100 per cent of cluster members), but they do represent only a small number of star formation sites that are biased in some way, i.e., they are unusually long-lived compared to most birth clusters. So, there is a great deal of information to be found in investigating the abundance behaviour of a large number of binary systems or co-moving stars (presumably co-natal systems in which 100 per cent of stars can be resolved, unlike compact, dense star clusters).

Since there are so many more of these pairs than star clusters, they should provide a diverse sampling of the various processes that are sources of noise in the chemical tagging signal --- for example, diffusion \citep{Dotter2017}; the possible abundance signature of exoplanets \citep{Melendez2009}; binary interactions \citep[e.g.,][]{Hansen2015}; and any underlying abundance inhomogeneity in star-forming environments \citep{Feng2014,Liu2016}. Any inference about chemical homogeneity based on data from a single cluster will depend on that cluster's particular formation environment and history. We cannot observe that environment directly, and it is not necessarily appropriate to apply the inference from one star-forming site to the entire disc, with its wide range of star-forming conditions across the history of the Milky Way. Even with a potentially significant false positive rate, the large sample of co-moving pairs reported by \citetalias{Oh2016} represents a large number of star formation sites, and they will give a more general picture of both the intrinsic chemical homogeneity of co-natal stars and the various processes that disturb that homogeneity.

With the release of \textit{Gaia} DR2, over 92 per cent of GALAH targets have proper motions errors less than 0.15~mas\,yr$^{-1}$ and parallaxes with a precision better than 0.1~mas. Nearly all of the dwarfs and many of the giants in the GALAH sample are expected to have precise parallax measurements from \textit{Gaia} DR2. This will make it possible to more thoroughly investigate dissolving but still kinematically related clusters and looser star formation sites. This will be aided by cases where we can determine ages with greater precision than the ages in this study, such as turn-off stars, massive main-sequence stars, or stars with spectroscopic age indicators \citep[e.g.,][]{Martig2016}, since stars formed at the same place and time must have the same age.

We are on the cusp of a dramatic change in our ability to understand the chemodynamical history of the Milky Way Galaxy. With GALAH and other surveys delivering chemical information and \textit{Gaia} delivering dynamical information, we will for the first time gain a statistically significant population of stars that we can chemically and dynamically tag together. 

\section*{Acknowledgements}
The GALAH survey is based on observations made at the Australian Astronomical Observatory, under programmes A/2013B/13, A/2014A/25, A/2015A/19, A/2017A/18. We acknowledge the traditional owners of the land on which the AAT stands, the Gamilaraay people, and pay our respects to elders past and present.

The following software and programming languages made this research possible: \textsc{configure} \citep{Miszalski2006}; \textsc{iraf} \citep{Tody1986,Tody1993}; Python (versions 3.6); \textsc{astropy} \citep[version 3.0;][]{Robitaille2013,TheAstropyCollaboration2018}, a community-developed core Python package for Astronomy; \textsc{pandas} \citep[version 0.20.2;][]{McKinney2010}; \textsc{topcat} \citep[version 4.4;][]{Taylor2005}; \textsc{galpy} \citep[version 1.3;][]{Bovy2015}.

This work has made use of data from the European Space Agency (ESA) mission {\it Gaia} (\url{https://www.cosmos.esa.int/gaia}), processed by the {\it Gaia} Data Processing and Analysis Consortium (DPAC, \url{https://www.cosmos.esa.int/web/gaia/dpac/consortium}). Funding for the DPAC has been provided by national institutions, in particular the institutions participating in the {\it Gaia} Multilateral Agreement.

Parts of this research were conducted by the Australian Research Council Centre of Excellence for All Sky Astrophysics in 3 Dimensions (ASTRO 3D), through project number CE170100013. SLM acknowledges support from the Australian Research Council through grant DE140100598. SB and KL acknowledge funds from the Alexander von Humboldt Foundation in the framework of the Sofja Kovalevskaja Award endowed by the Federal Ministry of Education and Research. LD gratefully acknowledges a scholarship from Zonta International District 24 and support from ARC grant DP160103747. KL acknowledges funds from the Swedish Research Council (Grant nr. 2015-00415\_3) and Marie Sklodowska Curie Actions (Cofund Project INCA 600398). TZ acknowledge the financial  support  from  the  Slovenian  Research  Agency  (research core funding No. P1-0188). A.~R.~C. acknowledges support through the Australian Research Council through grant DP160100637. LD, KF and Y-ST are grateful for support from Australian Research Council grant DP160103747.






\bsp    
\label{lastpage}
\end{document}